\documentclass[aps,11pt,preprintnumbers,nofootinbib,floatfix]{revtex4}

\usepackage{graphicx}
\usepackage{epsfig}
\usepackage{dcolumn}
\usepackage{subfigure}
\usepackage{multirow}

\def\be{\begin{equation}}
\def\ee{\end{equation}}
\def\bea{\begin{eqnarray}}
\def\eea{\end{eqnarray}}

\def\no{\nonumber}
\def\La{\Lambda}

\begin{document}

\title{Thermodynamic properties of black holes in de Sitter space}


\author{Huai-Fan Li$^{a,b}$, Meng-Sen Ma$^{a,b}$\footnote{Email: mengsenma@gmail.com}, Ya-Qin Ma$^{c}$}

\medskip

\affiliation{\footnotesize$^a$Department of Physics, Shanxi Datong
University,  Datong 037009, China\\
\footnotesize$^b$Institute of Theoretical Physics, Shanxi Datong
University, Datong 037009, China\\
\footnotesize$^c$Medical College, Shanxi Datong University, Datong 037009, China}


\begin{abstract}%
We study the thermodynamic properties of Schwarzschild-de Sitter (SdS) black hole and  Reissner-Nordstr\"{o}m-de Sitter (RNdS) black hole in the view of global and effective thermodynamic quantities.
Making use of the effective first law of thermodynamics, we can derive the effective thermodynamic quantities of de Sitter black holes. It is found that these effective thermodynamic quantities also satisfy Smarr-like formula.
Especially, the effective temperatures are nonzero in the Nariai limit.
By calculating heat capacity and Gibbs free energy, we find SdS black hole is always thermodynamically stable and RNdS black hole may undergoes phase transition at some points.
\end{abstract}

\maketitle

\section{Introduction}
Recently, thermodynamic criticality and phase transition of black holes in anti-de Sitter (AdS) space have been studied extensively.
Since the Hawking-Page phase transition for Schwarzschild-AdS black hole was proposed, various phase structures for different black holes in AdS space have been found.
It is found that the charged AdS black hole may have similar phase transition and critical behavior to that of Van der Waals liquid/gas system\cite{Chamblin1,Chamblin2,Lemos,Wu}.
There are also extensive studies on the thermodynamics and $P-V$ criticality of several kinds of black holes in an extended phase space by taking the cosmological constant as pressure\cite{Kastor,Dolan,Mann1,Mann2,Wu2,LYX,Ma1}. Meanwhile, astronomical observations show that our Universe is accelerating expansion\cite{AE,SPerl,AGR}, which means that it should be asymptotical de Sitter one.
Thus, it is necessary to consider the creation, dynamical evolution and thermodynamic properties of black holes in de Sitter space\cite{Kastor:1993,Ross,Hawking:1996}.

The major obstacle to study the thermodynamics of black holes in de Sitter space is the existence of multi-horizons and different temperatures for these horizons\cite{Lin, Paddy, SS}.
Generally, all de Sitter black holes have the black hole event horizon and the cosmological horizon.
The different temperatures for the two horizons make the whole spacetime cannot be in thermodynamic equilibrium. The conventional route to cope with de Sitter black holes is to treat the black hole horizon and cosmological horizon as two independent thermodynamic systems. For instance, one can analyze one horizon and take another one as the boundary or separate the two horizons by a thermally opaque membrane or box\cite{AG,Sekiwa,Cai:1998,Huang,Saida-a,Kubiznak}. In this way, the two horizons can be considered as two independent thermodynamic system and have their respective first law of thermodynamics. Besides, one can also take a global view to construct the globally effective temperature and other effective thermodynamic quantities\cite{Urano, Ma:IJMPA,Ma:2014,Ma:AHEP}. In this approach, the observer could imagine himself lie between the region between the black hole horizon and the cosmological horizon. For such an observer a new effective first law for the whole de Sitter black hole can be constructed with the total entropy assumed to be the sum of the entropies of the black hole horizon and the cosmological horizon\cite{Kastor:1993,SB}. However, it should be noted that there is also other choices for the total entropy of de Sitter black holes. As mentioned above, the whole system is in fact in a non-equilibrium thermodynamic state. The total entropy may be not simply the sum of entropies of the black hole horizon and the cosmological horizon. Even when multiple horizons exist for de Sitter  black holes, the area law of entropy may be no longer established for the cosmological horizon\cite{Saida-a}.

Even though adopting the globally effective approach to deal with de Sitter black holes, there are still two different starting points.
One can first take the geometric volume between the two horizons as the thermodynamic volume of the system and then derive the effective pressure and other effective thermodynamic quantities from the effective first law. In this case, the effective first law can be written in the form $dM=\tilde{T}_{eff}dS-\tilde{P}_{eff}dV+\cdots$, where the ``$\cdots$" represents the contributions from other matter fields. This idea has been employed in \cite{Urano, Ma:IJMPA,Ma:2014,Ma:AHEP}. The other idea is to consider the cosmological constant as variable and relate it to the pressure and then derive the effective volume and other effective thermodynamic quantities based on it. The effective first law in this case has the form $dM=T_{eff}dS+V_{eff}dP+\cdots$. Obviously, in the two cases, the effective thermodynamic quantities are different. Especially, the same $M$ in the two first laws have different meanings. In the former, it means internal energy and in the latter, the enthalpy. In this paper we will take the latter idea to study the thermodynamic properties of some de Sitter black holes. To study the thermodynamic stabilities of de Sitter black holes by means of effective thermodynamic, we also draw lessons from the former works on non-equilibrium thermodynamics\cite{Nieuwenhuizen1,Nieuwenhuizen2,Nieuwenhuizen3}. Although these effective
thermodynamic quantities are derived by analogy, they may reflect the global thermodynamic properties of de Sitter black hole. According to $(M,~P,~S)$ and the effective thermodynamic quantities $(T_{eff},~ V_{eff})$, we derive the effective heat capacity and Gibbs free energy, by which we can analyze the thermodynamic stability of the de Sitter black holes in the global sense.

The paper is arranged as follows. In Section II, we study the thermodynamics of Schwarzschild-de Sitter black hole. In Section III we will study Reissner-Nordstr\"{o}m-de Sitter black hole and
analyze its thermodynamic stability. We  make some
concluding remarks in Section V.

\section{Thermodynamics of Schwarzschild-de Sitter (SdS) black hole}

For static, spherically symmetric spacetime, the metric ansatz is usually given by
\begin{equation}\label{metric}
ds^2=-f(r)dt^2+f(r)^{-1}dr^2 + r^2d\Omega^2,
\end{equation}
where
\be
f(r)=1-\frac{2 M}{r}-\frac{\Lambda  r^2}{3},
\ee
for SdS black hole.

If $0<9\Lambda M^2<1$, the two positive roots of $f(r)=0$ mark the positions of the horizons of SdS spacetime.
The smaller one, expressed as $r_{+}$, corresponds to the black hole event horizon.
And the larger one, $r_{c}$, corresponds to the cosmological horizon.

One can express the parameters $M,~\Lambda$ according to $r_{+}$ and $r_c$. They are
\be
M=\frac{x (x+1) r_c}{2 \left(x^2+x+1\right)}, \quad \Lambda=\frac{3}{\left(x^2+x+1\right) r_c^2},
\ee
where we have set $x=r_+/r_c$, thus $0\leq x \leq 1$.

The surface gravities of black hole horizon and the cosmological horizon are:
\be\label{ksds}
\kappa_{+}=\frac{1-\Lambda r_{+}^2}{2r_{+}}>0, \quad \kappa_{c}=\frac{1-\Lambda r_{c}^2}{2r_{c}}<0, \quad \text{and} \quad \kappa_{+}>|\kappa_{c}|,
\ee
because $r_{+}$ and $r_{c}$ satisfy the relation:
\be
2M<r_{+}<3M<\sqrt{\frac{1}{\La}}<r_{c}<\sqrt{\frac{3}{\La}}.
\ee
Temperatures of both the horizons are
\be
T_{+}=\frac{\kappa_{+}}{2\pi}, \quad T_{c}=-\frac{\kappa_{c}}{2\pi}.
\ee
Entropies for the two horizons are respectively:
\be
S_{+}=\pi r_{+}^2, \quad S_{c}=\pi r_{c}^2.
\ee
We consider the cosmological constant $\La$ as variable and relate it to the thermodynamic pressure: $P=-\La/8\pi$\cite{Dolan:PRD}.
It is found that the first laws of thermodynamics can be established on the two horizons:
\be\label{1st}
dM=\frac{\kappa_{+}}{2\pi}dS_{+}+V_{+}dP, \quad
dM=\frac{\kappa_{c}}{2\pi}dS_{c}+V_cdP,
\ee
where $V_{+}=\frac{4\pi r_{+}^3}{3}$ and $V_{c}=\frac{4\pi r_{c}^3}{3}$ are the conjugate volumes to the thermodynamic pressure.

Now let us take a global view on the thermodynamics of the SdS spacetime.
In our former works, we have studied the thermodynamics of SdS black holes based on the effective first law: $dM=\tilde{T}_{eff}dS-\tilde{P}_{eff}dV$
with $V=4\pi(r_{c}^3-r_{+}^3)/3$, the volume between the black hole horizon and the cosmological horizon and the total entropy $S=S_{+}+S_{c}$\cite{Zhao}.
In this framework, $\tilde{T}_{eff}=\left.\frac{\partial{M}}{\partial{S}}\right|_V$, is the effective temperature and $\tilde{P}_{eff}=-\left.\frac{\partial{M}}{\partial{V}}\right|_S$, is the effective pressure. Below we take another view.

One can introduce the effective thermodynamic quantities and construct effective thermodynamic law for the SdS spacetime:
\be
dM=\frac{\kappa_{eff}}{2\pi}dS+V_{eff}dP,
\ee
where $\kappa_{eff}$ is the effective surface gravity and  $V_{eff}$, the effective thermodynamic volume. We can derive the effective thermodynamic quantities\footnote{Here $\kappa_{eff}$ is negative. This can be seen from Eq.(\ref{ksds}). Physically acceptable temperature should be positive, so we take the absolute value. }:
\bea
T_{eff}=-\frac{\kappa_{eff}}{2\pi}&=&-\left.\frac{\partial{M}}{\partial{S}}\right|_P=\frac{(x+2) (2 x+1)}{4 \pi  (x+1) \left(x^2+x+1\right) r_c}=\frac{\Lambda  (x+2) (2 x+1)}{4 \sqrt{3} \pi  (x+1) \sqrt{\Lambda  \left(x^2+x+1\right)}}\label{TSDS} \\
V_{eff}&=&\left.\frac{\partial{M}}{\partial{P}}\right|_S=\frac{4 \pi  \left(x^4+3 x^3+3 x^2+3 x+1\right) r_c^3}{3 (x+1)}.\label{VSDS}
\eea
One can easily verify that these effective thermodynamic quantities also satisfy the Smarr-like formula:
\be
M=-2T_{eff}S-2V_{eff}P.
\ee
This relation hints the scaling behaviors of these effective thermodynamic variables. Obviously, when the cosmological horizon radius $r_c$ is scaled as $r_c \rightarrow \lambda r_c$,
other state variables are scaled as follows: $\Lambda \rightarrow \Lambda/\lambda^2, ~ M \rightarrow \lambda M$, $T_{eff} \rightarrow T_{eff}/\lambda$ and $ V_{eff} \rightarrow  \lambda^3 V_{eff}$. This issue has been carefully discussed in \cite{Saida-a,Saida-b}.

In this form, $M$ should be identified with the enthalpy of the whole SdS thermodynamic system. When $x\rightarrow 0$, the black hole horizon vanishes, SdS spacetime becomes pure dS space. In this case, one can easily find that
\be
T_{eff}=\frac{1}{2 \pi  r_c}, \qquad V_{eff}=\frac{4 \pi  r_c^3}{3}.
\ee
 These are exactly thermodynamic quantities of dS space. When $x\rightarrow 1$, the black hole horizon and the cosmological horizon \emph{apparently} coincide, but the volume between them is not zero. The nonzero volume indicates that in this limit the region between the two horizons does not shrink into zero. The two horizons are not really coincident. This is usually called Nariai limit\cite{Nariai,Ginsparg}. In this case, $r_{+}=r_{c}=r_{0}=\sqrt{1/\La}$, thus $T_{+}=T_{c}=0$.

 However, as Bousso and Hawking had pointed out\cite{Hawking:1996}, for SdS black hole the normalization constant $\gamma_{t}$ of timelike Killing vector field $K=\gamma_{t}\frac{\partial}{\partial t}$ cannot set to be $\gamma_{t}=1$ as usual, but should be taken as $\gamma_{t}=\frac{1}{\sqrt{f(r_g)}}$ with $r_g$ a reference point. In this way, the Bousso-Hawking temperature can be calculated along with the new normalization as $T^{BH}_{+}=\frac{f'(r_{+})}{4\pi \sqrt{f(r_g)}}$ and $T^{BH}_{c}=\frac{f'(r_{c})}{4\pi \sqrt{f(r_g)}}$. One can easily find that the Bousso-Hawking temperature is not zero in the Nariai limit, but $T^{BH}_{+}=T^{BH}_{c}=\frac{\sqrt{\Lambda}}{2\pi}$.

 According to Eqs.(\ref{TSDS}), (\ref{VSDS}), in the Nariai limit  the effective temperature and effective volume turn into
\be
T_{eff}=\frac{3}{8 \pi  r_0}=\frac{3\sqrt{\Lambda}}{8\pi}, \qquad V_{eff}=\frac{22 \pi  r_0^3}{3}.
\ee
Clearly, these results are consistent with that of Bousso and Hawking, although the coefficient is a little different.

\begin{figure}[!htbp]
\includegraphics[width=3in]{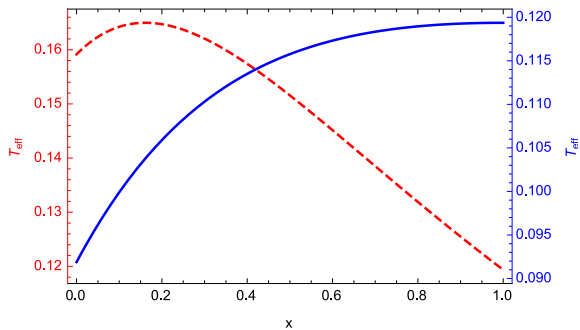}
\caption{$T_{eff}$ as function of $x$. The solid (blue) curve depicts $T_{eff}(x,\La)$ with $\La=1$. The dashed (red) curve depicts $T_{eff}(x,r_{c})$ with $r_{c}=1$.}
\label{Tsds}
\end{figure}

In Fig.\ref{Tsds}, we have depicted the effective temperature $T_{eff}$. It is shown that $T_{eff}$  monotonically increases as the $x$ increases when $\La$ is fixed. In the Nariai limit, it reaches the maximum. When the cosmological horizon $r_c$ is fixed, the effective temperature will first increase as the black hole horizon $r_{+}$ increases, until the maximum, and then decrease. In the Nariai limit, it reaches the minimum.
To understand the thermodynamic stability of the SdS thermodynamic system, we can calculate the heat capacity by means of the effective thermodynamic quantities.
\be
C_P=\left.\frac{\partial{M}}{\partial{T_{eff}}}\right|_P=\frac{2 \pi  (x+1)^2 (x+2) (2 x+1) r_c^2}{4 x^2+7 x+4}.
\ee
Obviously, the heat capacity at constant pressure is always positive. Thus, we conclude that the global SdS  system is thermodynamically stable and no phase transition arises.

\section{Thermodynamics of Reissner-Nordstr\"{o}m-de Sitter (RNdS) black hole }





For the RNdS spacetime, the metric ansatz is also given by Eq.(\ref{metric}), but with
\be
f(r)=1-\frac{2 M}{r}+\frac{Q^2}{r^2}-\frac{\Lambda  r^2}{3}.
\ee
There are three positive real roots for $f(r)=0$. The smallest one $r_{-}$ is the inner/Cauchy horizon,
the intermediate one $r_{+}$ is the event horizon of black hole and the largest one is the cosmological horizon.
According to $r_{+}, ~r_{c}$ and $Q$, we can express $M,~ \La$ as:
\be
M=\frac{(x+1) \left(x^2 r_c^2+Q^2 x^2+Q^2\right)}{2 x \left(x^2+x+1\right) r_c}, \quad \La = \frac{3 \left(x r_c^2-Q^2\right)}{x \left(x^2+x+1\right) r_c^4}.
\ee
When the black hole horizon and the cosmological horizon are viewed as independent each other, there are also respective first laws of thermodynamics:
\be\label{1st}
dM=\frac{\kappa_{+}}{2\pi}dS_{+}+\Phi_{+}dQ+V_{+}dP, \quad
dM=\frac{\kappa_{c}}{2\pi}dS_{c}+\Phi_{c}dQ+V_cdP,
\ee
where $\Phi_{+}$ and $\Phi_{c}$ are electric potentials corresponding to the two horizons.

Considering the connection between the black hole horizon and the cosmological horizon, we can derive the effective thermodynamic quantities and corresponding first law of black hole thermodynamics:
\be\label{eff1st}
dM = \frac{\kappa_{eff}}{2\pi}dS + {\Phi _{eff}}dQ + {V_{eff}}dP.
\ee
The effective surface gravity $\kappa_{eff}$, the effective electric potential $\Phi_{eff}$ and the effective thermodynamic volume $V_{eff}$ are respectively:
\bea
\frac{\kappa_{eff}}{2\pi}&=&\left.\frac{\partial{M}}{\partial{S}}\right|_{Q,P}\no \\
       &=&-\frac{\left[Q^2 \left(x^2+2 x+3\right)-x (x+2) r_c^2\right] \left[Q^2 \left(3 x^2+2 x+1\right)-x^2 (2 x+1) r_c^2\right]}{4 \pi  x (x+1) \left(x^2+x+1\right) r_c^3 \left[x^2 r_c^2-Q^2 (x+1)^2\right]},\\
\Phi_{eff}&=&\left.\frac{\partial{M}}{\partial{Q}}\right|_{S,P}=-\frac{Q \left[x^2 r_c^2+Q^2 \left(x^2+1\right)\right]}{(x+1) r_c \left[x^2 r_c^2-Q^2 (x+1)^2\right]},\\
V_{eff}&=&\left.\frac{\partial{M}}{\partial{P}}\right|_{S,Q}\no \\
        &=&\frac{4 \pi  r_c^3 \left[x^2 \left(x^4+3 x^3+3 x^2+3 x+1\right) r_c^2-Q^2 \left(x^6+3 x^5+6 x^4+6 x^3+6 x^2+3 x+1\right)\right]}{3 (x+1) \left[x^2 r_c^2-Q^2 (x+1)^2\right]}.\label{VRN}
\eea
When $Q=0$, RNdS black hole returns back to SdS black hole. Correspondingly, the above quantities indeed degenerate to Eqs.(\ref{TSDS}) and (\ref{VSDS}), only if we define the effective temperature of the RNdS black hole to be:
\be\label{TRNDS}
T_{eff}=\frac{|\kappa_{eff}|}{2\pi}.
\ee
Also, these effective thermodynamic quantities satisfy the Smarr-like formula:
\be
M=-2T_{eff}S+\Phi_{eff}Q-2V_{eff}P,
\ee
which also reflects the scaling behaviors of the thermodynamic state variables.

For the RNdS black hole, it is meaningless to discuss the $x \rightarrow 0$ limit. In this limit, the RNdS spacetime will not return to the SdS or dS spacetime.
In the $x \rightarrow 1$ limit, the charged Nariai black hole is obtained. In this case, $r_{+}=r_{c}=\rho$.  $M$ and $\La$  have simple forms:
\be
M=\frac{\rho ^2+2 Q^2}{3 \rho }, \quad \La=\frac{\rho ^2-Q^2}{\rho ^4}.
\ee
The inner horizon lies at
\be
r_{-}=b=\frac{\sqrt{-\Lambda  \left(2 \Lambda  \rho ^2-3\right)}}{\Lambda }-\rho=\frac{\rho  \left(-\rho ^2+Q^2+\sqrt{\rho ^4-2 Q^4+\rho ^2 Q^2}\right)}{\rho ^2-Q^2}<\rho.
\ee
The effective temperature $T_{eff}$ in the charged Nariai case, is
\be
T_{eff}=\frac{\left(b^2+2 b \rho -3 \rho ^2\right)^2}{8 \pi  \rho  \left(\rho ^2 -b^2 - 2 b \rho\right) \left(b^2+2 b \rho +3 \rho ^2\right)}\neq 0.
\ee
In the ultracold case, namely the three horizons coincide, one can easily find that $T_{eff}=0$.

Now it is time to discuss the thermodynamic properties of RNdS black hole. Without loss of generality, we will take the choice $r_c=1$ below. It is found that $T_{eff}$ is always negative when $Q^2>1/2$.
A negative temperature does not make any sense in black hole thermodynamics. Thus, below we will study the thermodynamic properties of RNdS black hole with $0\leq Q^2 \leq 1/2$.

\begin{figure}[!h]
\centering\includegraphics[width=3in]{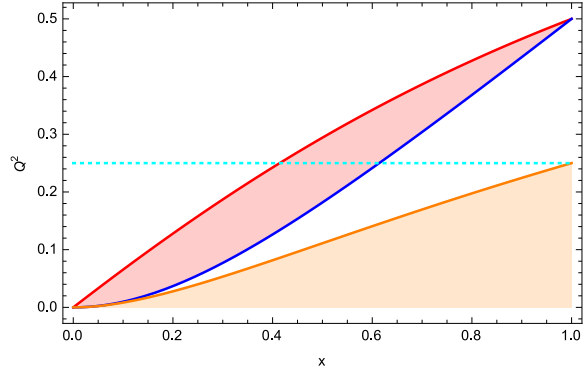}
\caption{The relations between $x$ and $Q^2$. Only in the shadow regions $T_{eff}$ is positive. The dotted line corresponds to $Q^2=1/4$.}
\label{Q2x}
\end{figure}

\begin{figure}[!h]
\center{ \subfigure[$Q=0.2$]{ \label{T-a}
\includegraphics[width=2.5in]{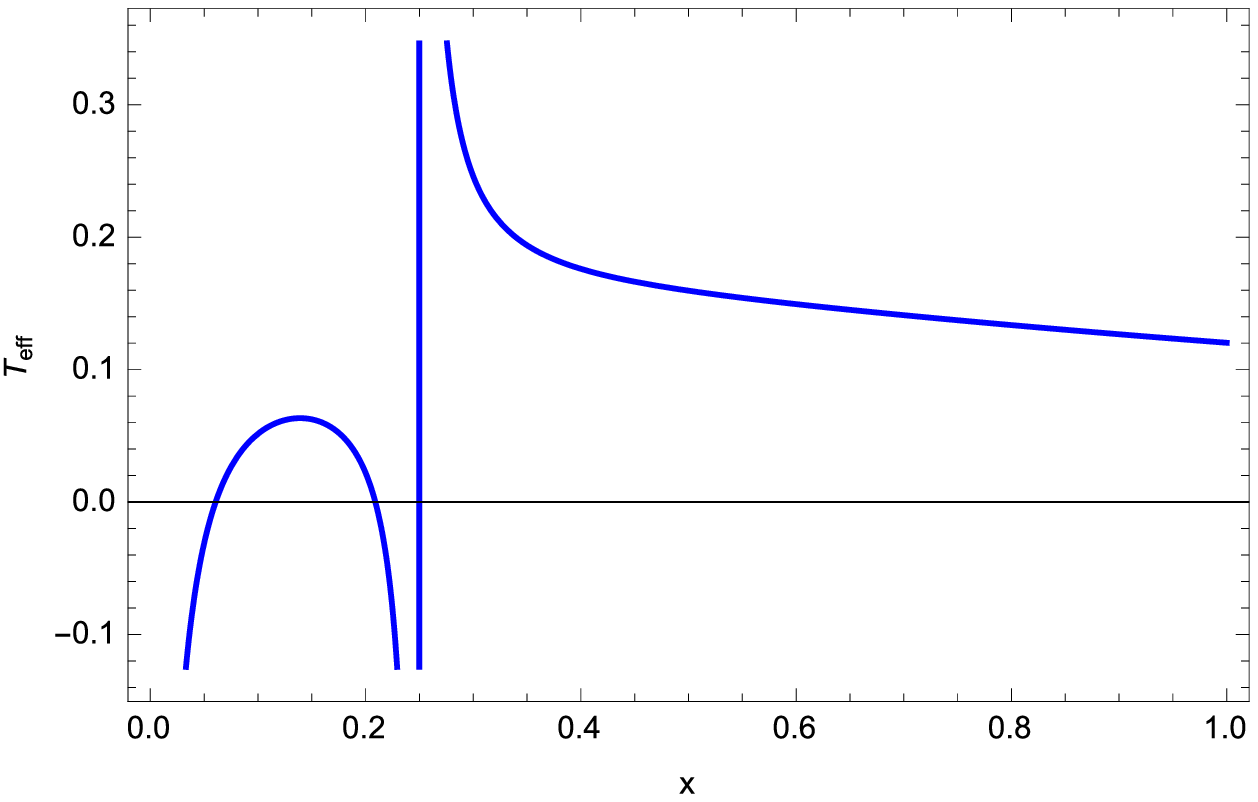} \hspace{0.5cm}}
\subfigure[$Q=0.55$]{ \label{T-b}
\includegraphics[width=2.5in]{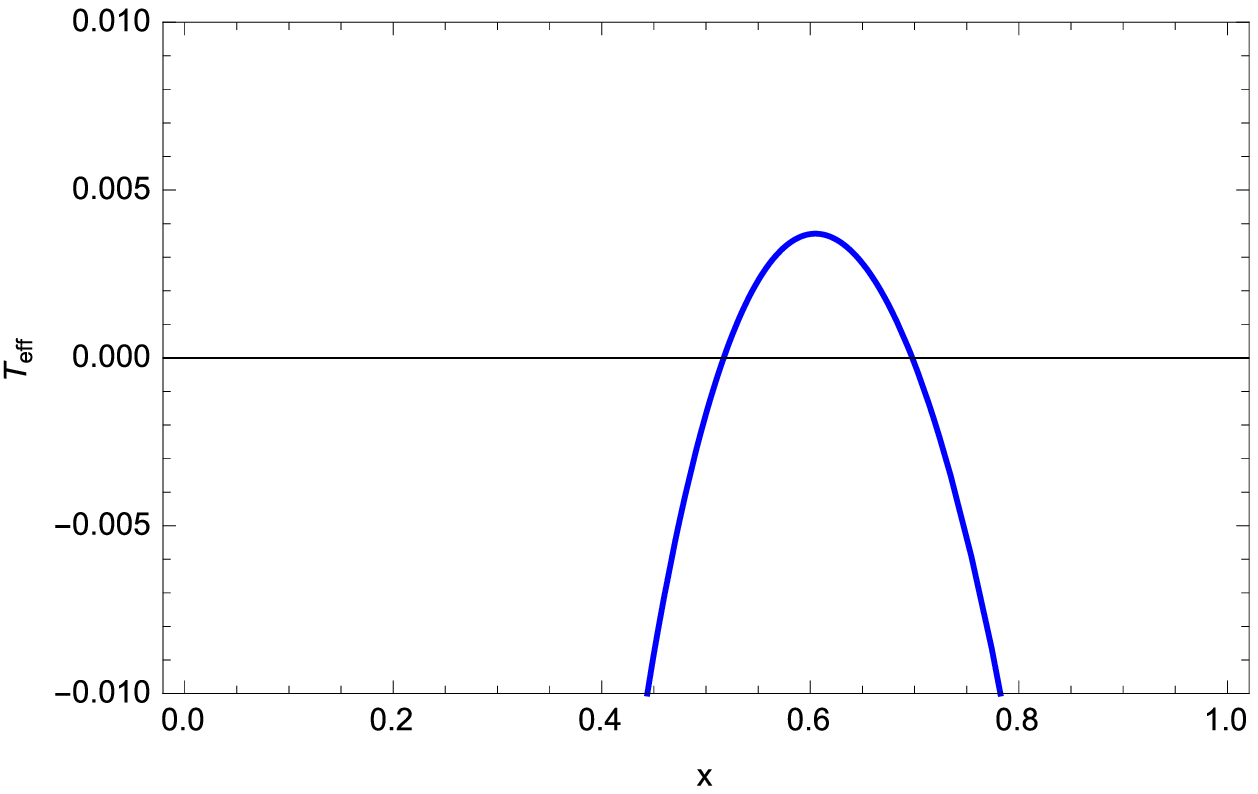}}
\caption{The effective temperature $T_{eff}$ as functions of $x$. In (a), the two zero-temperature points lie at $x=0.06$ and $x=0.21$. The divergent point lies at $x=0.25$. In (b), the two zero-temperature points are
$x=0.52$ and $x=0.7$. }}
\label{TRN}
\end{figure}

To require $T_{eff}\geq 0$, there are some restrictions on $Q$ and $x$, which are shown in Fig.\ref{Q2x}.
For fixed $x$, there are always two regions in which the effective temperature $T_{eff}$ is positive. While for fixed $Q$, there will be one or two regions in which $T_{eff}$ is positive,  depending on the values of $Q$. As is shown in Fig.\ref{Q2x}, there is one region when $Q^2>1/4$ and  two regions when $Q^2 \leq 1/4$. Therefore, $T_{eff}$ has different behaviors for $Q^2 \in [0,1/4]$ and $Q^2 \in (1/4,1/2]$.
In Fig.\ref{TRN}, we plot the $T_{eff}$ for $Q=0.2$ and $Q=0.55$, respectively. It should be noted that the two positive regions are disconnected, as is shown in Fig.\ref{T-a}. The black hole cannot transit through a negative temperature region. Thus, the RNdS black hole can only stay in one of the two regions and will be always in that region.

\begin{figure}[!h]
\center{ \subfigure[$Q=0.2$]{ \label{C-a}
\includegraphics[width=2.5in]{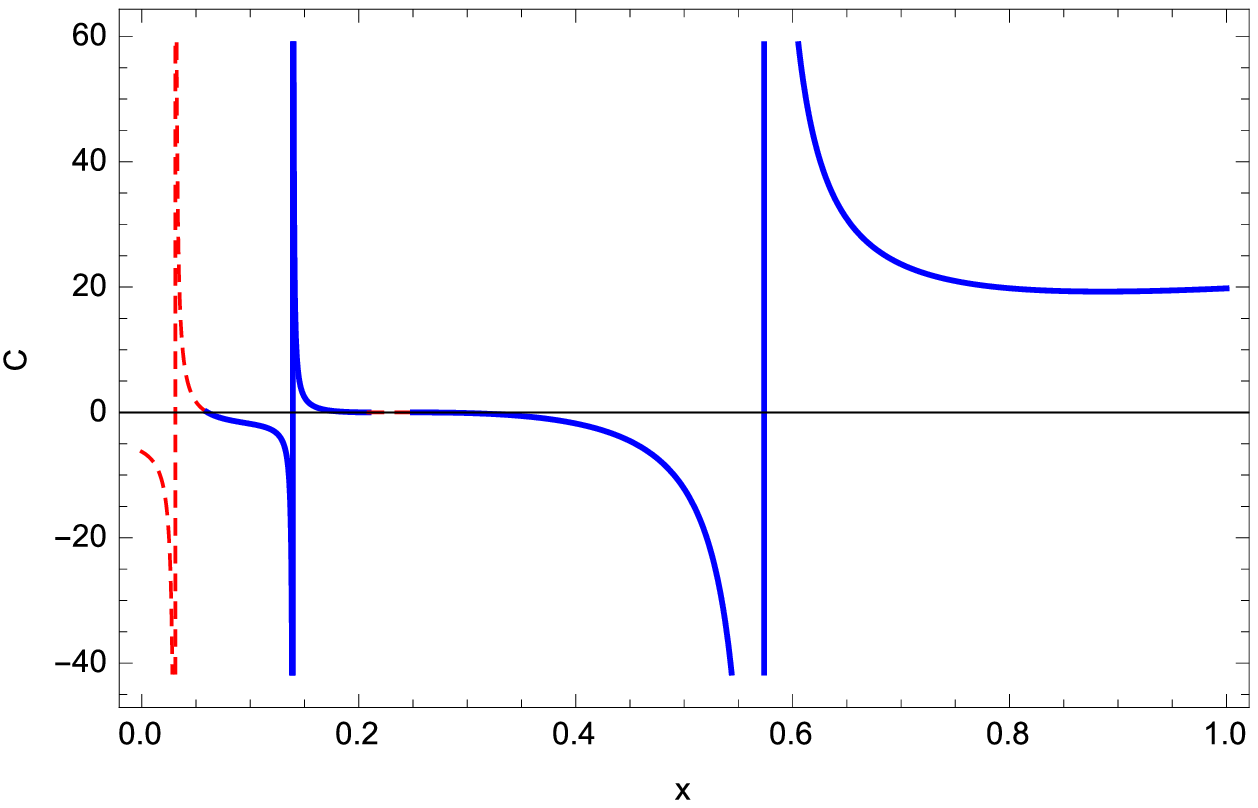} \hspace{0.5cm}}
\subfigure[$Q=0.55$]{ \label{C-b}
\includegraphics[width=2.5in]{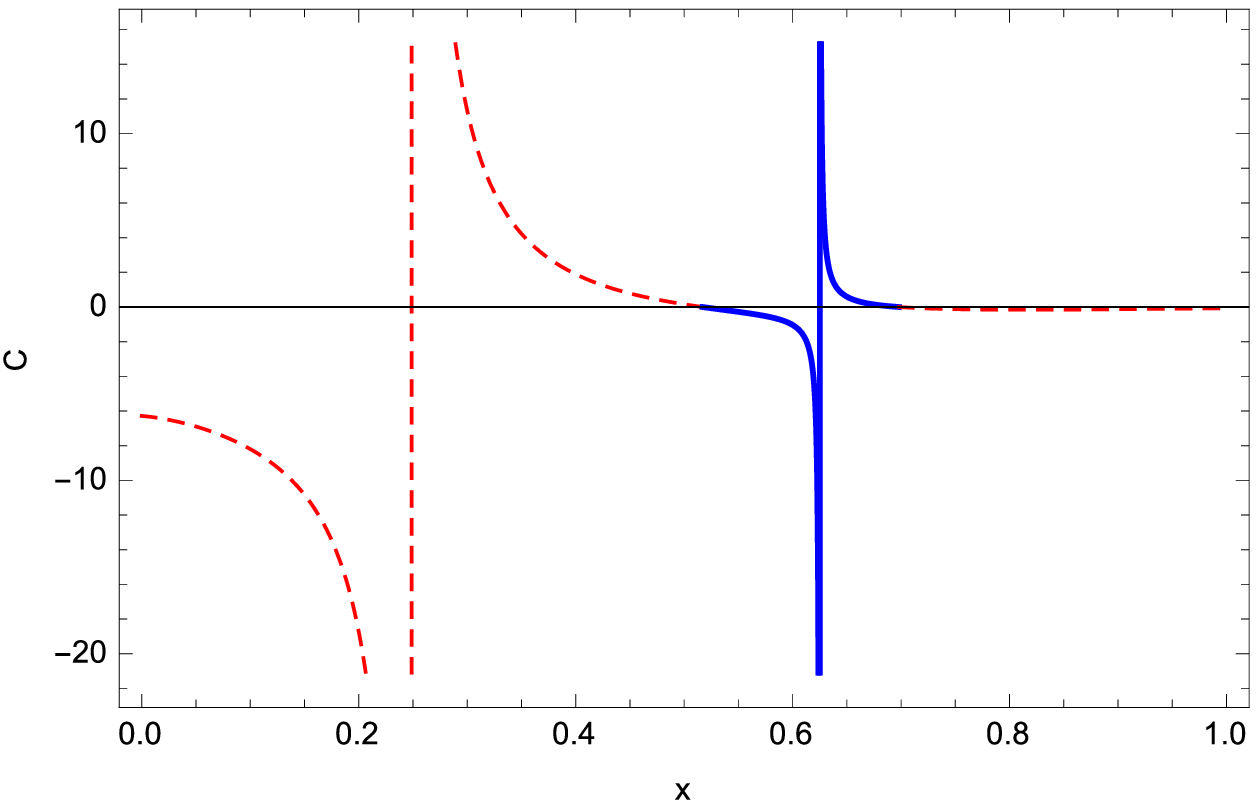}}
\caption{The heat capacity $C$ as functions of $x$. The solid (blue) curves correspond to positive-temperature regions. In (a), the two divergent points are respectively $x=0.14$ and $x=0.57$. In (b), the heat capacity diverges at $x=0.62$.}}
\label{CRN}
\end{figure}

The heat capacity can be defined as
\be
C=\left.\frac{\partial{M}}{\partial{T_{eff}}}\right|_{Q,P}.
\ee
Its complete expression is very lengthy. We will put it in the Appendix. The behaviors of the heat capacity $C$ for $Q=0.2$ and $Q=0.55$ are depicted in Fig.4. The red dashed curves, which correspond to negative effective temperature regions, are meaningless. The left divergent point in Fig.\ref{C-a} and the divergent point in Fig.\ref{C-b} both correspond to the points where $T_{eff}$ has local maximum values. However, another unexpected divergence for $C$ arises in Fig.\ref{C-a}. The heat capacity can be positive in each positive-temperature region. In Fig.\ref{C-b}, the larger black hole with positive heat capacity should be thermodynamic stable. In the two disconnected positive-temperature regions in Fig.\ref{C-a}, both the black holes with larger $x$ have positive heat capacity. Therefore the two cases are both thermodynamically stable. However, only by heat capacity we cannot judge which region is more thermodynamically preferred.

To discuss the global stability of the RNdS black hole,
we need to calculate the Gibbs free energy. Because $M$  is now viewed as enthalpy, the Gibbs free energy is defined as
\be
G=M-T_{eff}S.
\ee
Also, its complete expression is given in the Appendix.
Generally, states with smaller $G$ are more thermodynamically stable.

\begin{figure}[!h]
\center{ \subfigure[$Q=0.2$]{ \label{G-a}
\includegraphics[width=4cm]{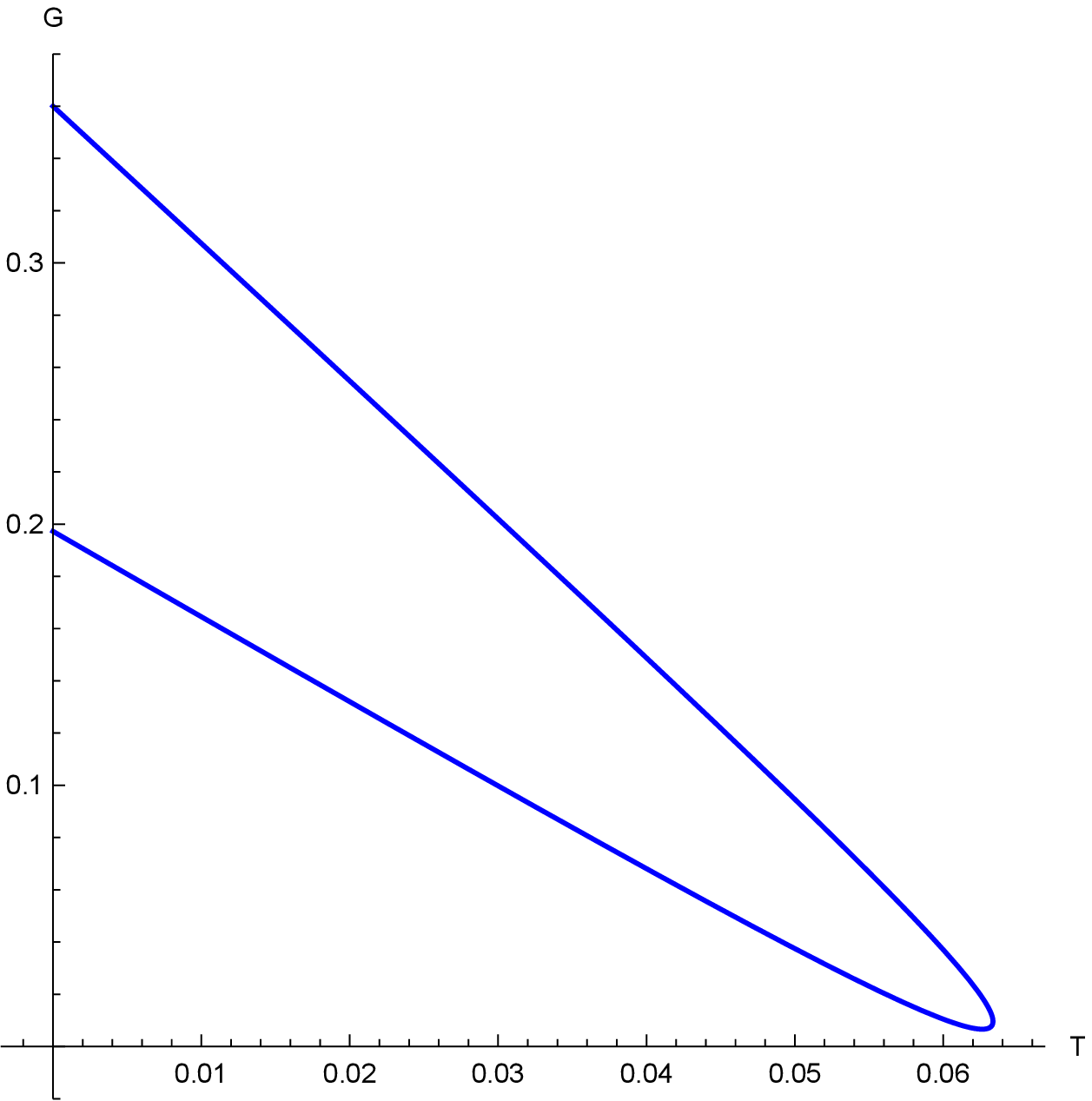} \hspace{0.5cm}}
\subfigure[$Q=0.2$]{ \label{G-b}
\includegraphics[width=4cm]{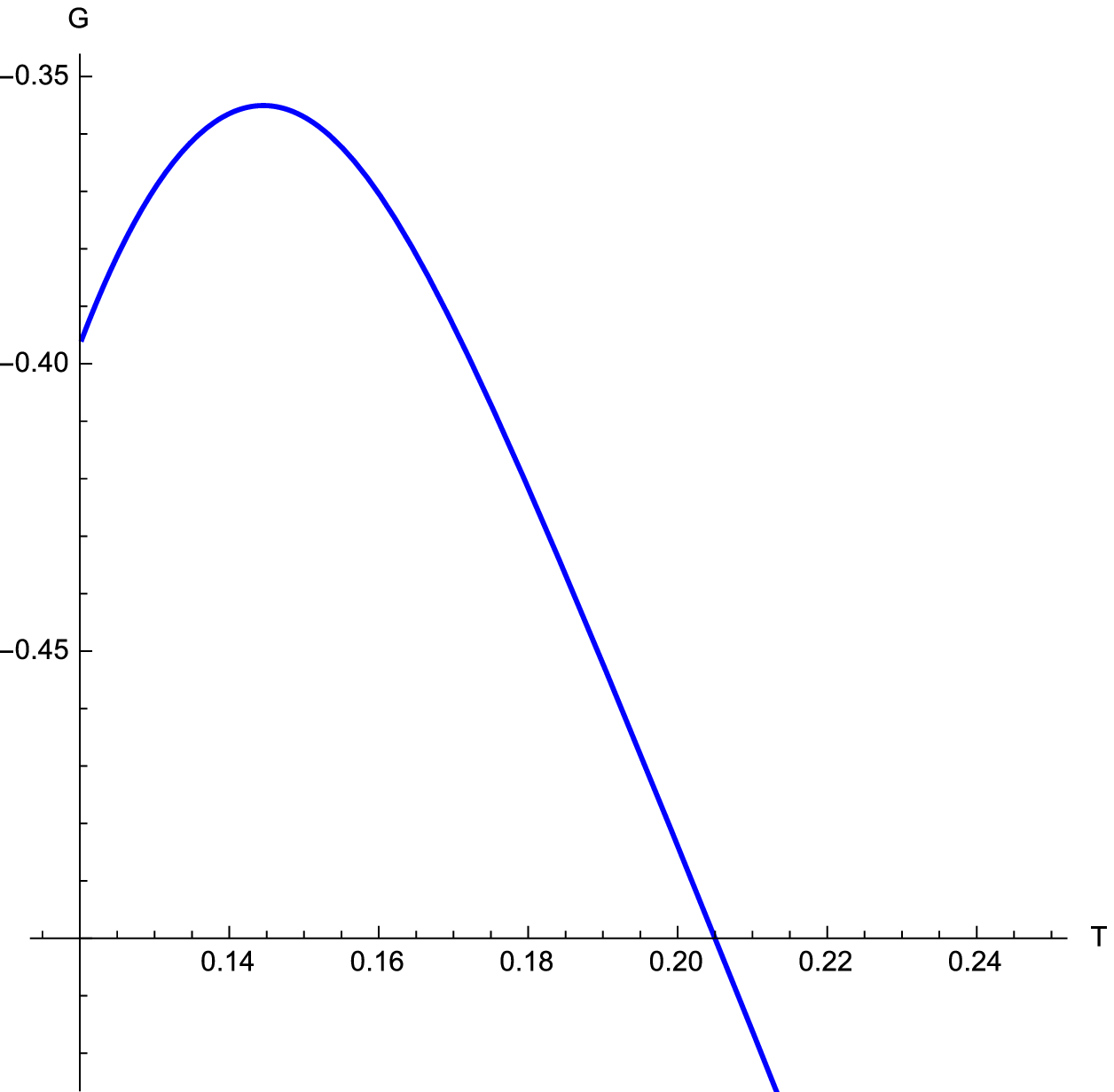}\hspace{0.5cm}}
\subfigure[$Q=0.55$]{ \label{G-c}
\includegraphics[width=4cm]{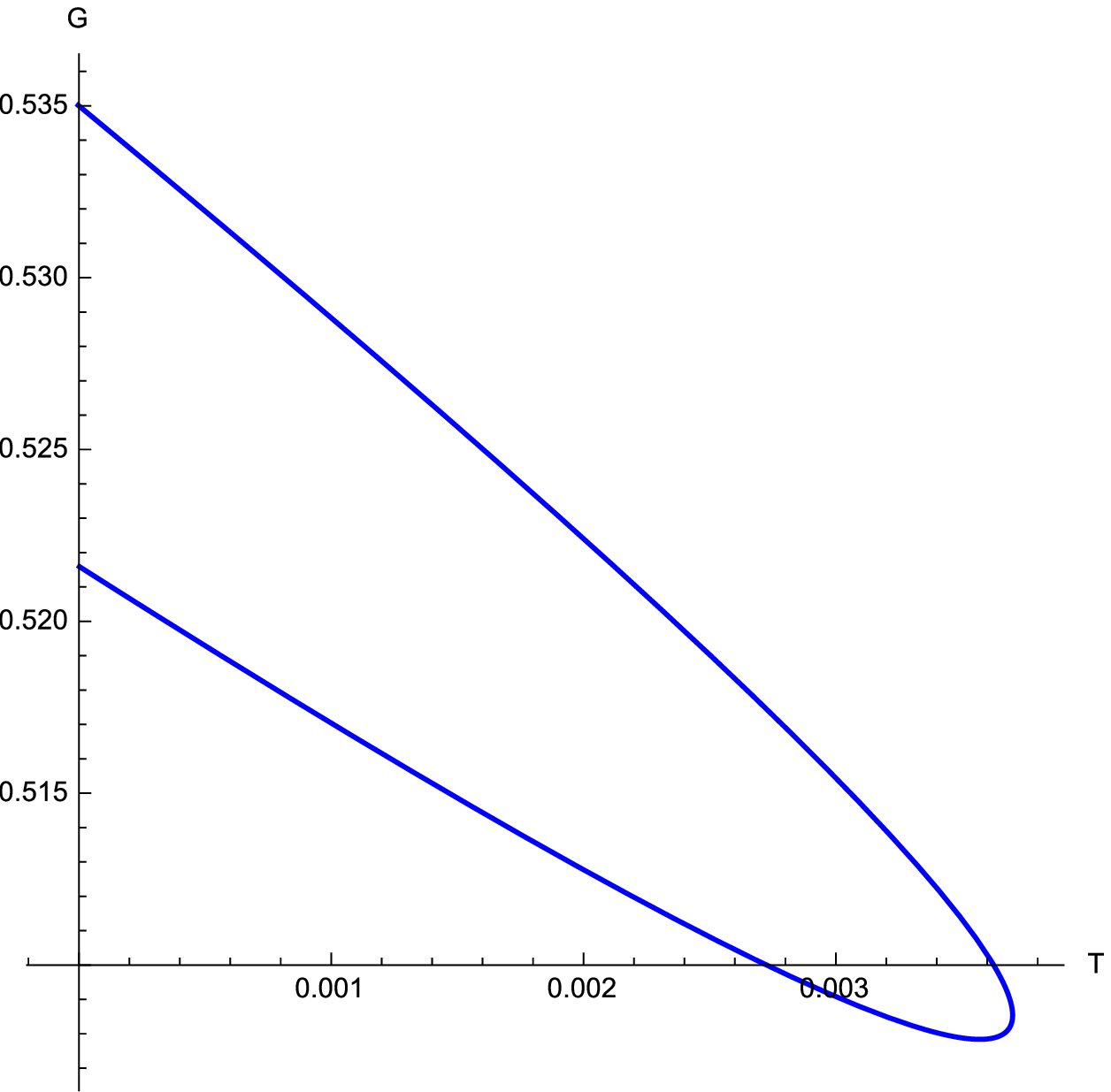}}
\caption{The Gibbs free energy as functions of $T_{eff}$. (a) corresponds to the left positive-temperature region in Fig.\ref{T-a}, (b) corresponds to the right positive-temperature region in Fig.\ref{T-a} and (c) corresponds to the positive-temperature region in Fig.\ref{T-b}. }}
\label{GRN}
\end{figure}

As is shown in Fig.\ref{G-a} and Fig.\ref{G-c}, the lower branches correspond to black holes with positive heat capacity,which are indeed thermodynamically stable. However, comparing Fig.\ref{G-a} with Fig.\ref{G-b}, we can find that in former case Gibbs free energy is always positive , while it is always negative in the latter case. Therefore, we conclude that the right positive-temperature region is more thermodynamically preferred because of the lower $G$.

\section{Conclusion}

In this paper, we have presented the globally thermodynamic properties including thermodynamic stability and phase transition of SdS black hole and RNdS black hole.
For de Sitter black holes, the black hole horizon and the cosmological horizon can both be treated as a thermodynamic system. However, the temperatures on the two horizons are usually different. Thus the two horizons cannot be in thermodynamic equilibrium, except for the Nariai or lukewarm case. Considering the two horizons share the same parameters $M,~Q,~\La$, there should be some connections between the thermodynamic quantities of the two horizons.
Thus, we take a global view to deal with the de Sitter black holes. After treating the cosmological constant as the thermodynamic pressure, we can construct the effective first law of thermodynamics, from which the effective thermodynamic quantities can be derived. It is interesting to find that the effective temperatures for SdS black hole and RNdS black hole tend to a nonzero value in the Nariai limit. This result agrees with the result of Bousso and Hawking.

For SdS black hole, the effective temperature is always positive. For RNdS black hole, the effective temperature has two different behaviors depending to the values of $Q$. When $Q^2 \leq 1/4$, the effective temperature can be positive in two disconnected regions. When $1/4<Q^2 \leq 1/2$, the effective temperature is positive in one region. When $Q^2>1/2$, the effective temperature is always negative.
We also discussed the locally thermodynamic stability by calculating heat capacity. It is shown that the heat capacity is always positive for SdS black hole, which mean SdS black hole is always thermodynamic stable. For RNdS black hole, heat capacity can have positive values in each positive-temperature regions when $1/4<Q^2 \leq 1/2$. However, according to Gibbs free energy we find that RNdS black hole is more thermodynamically preferred in the right positive-temperature region.

\section*{Acknowledgment}

The author also would like to thank Professor Ren Zhao for useful discussion. This work is supported in part by the Young Scientists Fund of the National Natural Science Foundation
of China (Grant Nos.11205097, 11605107),the Natural Science Foundation for Young Scientists of
Shanxi Province, China (Grant No.2012021003-4), and by the Doctoral Sustentation Fund of Shanxi Datong
University (2011-B-03).


%

\appendix

\section{Complete expressions of $C$ and $G$}

Here we show the complete forms of heat capacity and the Gibbs free energy ($r_c=1$):
\be
C=\frac{2 \pi  (x+1)^2}{A(x)} \left[x^2-Q^2 (x+1)^2\right]^2 \left[x^2+2x-Q^2 \left(x^2+2 x+3\right)\right] \left[2 x^3+x^2-Q^2 \left(3 x^2+2 x+1\right)\right],
\ee
where
\bea
A(x)&=&-3 Q^8 (x+1)^2 \left(x^6+8 x^5+11 x^4+16 x^3+11 x^2+8 x+1\right) \no \\
     &+& 2 Q^6 x \left(2 x^8+23 x^7+59 x^6+81 x^5+90 x^4+81 x^3+59 x^2+23 x+2\right)\no \\
     &-& 6 Q^4 x^3 \left(2 x^6+2 x^5-7 x^4-12 x^3-7 x^2+2 x+2\right) \no \\
     &-&6 Q^2 x^5 \left(2 x^4+9 x^3+13 x^2+9 x+2\right)+x^7 \left(4 x^2+7 x+4\right).
\eea
 \bea
 G&=&-\frac{1}{B(x)}\left[Q^4 \left(5 x^6+16 x^5+31 x^4+32 x^3+31 x^2+16 x+5\right) \right. \no \\
   &-& \left.2 Q^2 x \left(x^6+4 x^5+7 x^4+4 x^3+7 x^2+4 x+1\right)+x^3 \left(2 x^4+3 x^3+3 x+2\right)\right],
 \eea
where
\be
B(x)=4 x (x+1) \left(x^2+x+1\right) \left(x^2-Q^2 (x+1)^2\right).
\ee

\end{document}